\documentclass[pre,twocolumn,10pt,aps,longbibliography]{revtex4-1}

\usepackage{ulem}
\usepackage{amsmath,amsfonts,amssymb}
\usepackage{amssymb}
\usepackage{amsthm}
\usepackage[dvipdf]{color}
\usepackage{graphicx}
\usepackage{dcolumn}
\usepackage{bm} 
\usepackage[rightcaption]{sidecap}

\begin{document}

\title{Topological Boundary Modes in Nonlinear Dynamics with Chiral Symmetry}

\author{Di Zhou}
\email{dizhou@bit.edu.cn}
\affiliation{Key Lab of Advanced Optoelectronic Quantum Architecture and Measurement (MOE) and School of Physics, Beijing Institute of Technology, Beijing 100081, China}

\affiliation{Department of Physics, University of Illinois, Urbana-Champaign, 61801, USA}

\begin{abstract}
Particle-hole symmetry and chiral symmetry play a pivotal role in multiple areas of physics, yet they remain unstudied in systems with nonlinear interactions whose nonlinear normal modes do not exhibit $\textbf{U}(1)$-gauge symmetry. In this work, we establish particle-hole symmetry and chiral symmetry in such systems. Chiral symmetry ensures the quantization of the Berry phase of nonlinear normal modes and categorizes the topological phases of nonlinear dynamics. We show topologically protected static boundary modes in chiral-symmetric nonlinear systems. Our theoretical framework extends particle-hole and chiral symmetries to nonlinear dynamics, whose nonlinear modes do not necessarily yield $\textbf{U}(1)$-gauge symmetry. 
\end{abstract}

\maketitle

\section{Introduction}

Non-spatial symmetries govern the fundamental principles of physics in multiple areas. In high-energy physics, the combination of particle-hole, parity, and time-reversal symmetries dictate the existence of anti-particles~\cite{RevModPhys.71.373}. In quantum mechanics, time-reversal symmetry demands that eigenstates with half-integer spins must be doubly degenerate, which is known as the Kramers' degeneracy~\cite{Zhang2011RMP, Niu2010RMP}. In soft matter and engineering physics, chiral symmetry reveals the chiral image of mechanical floppy modes and states of self-stress~\cite{Lubensky2015phonons, Kane2014NP, Zhou2018PRL, VitelliPRX}, governing the mechanical failure and stability~\cite{Sun2020PRL, paulose2015topological, Zhou2019PRX, Chen2023IM}, respectively. Additionally, in condensed-matter physics, time-reversal, particle-hole, and chiral symmetries classify topological phases of matter in a ``ten-fold" way~\cite{RevModPhys.88.035005, schnyder2008classification, Kane2010RMP, Su1979PRL}. This classification enables fundamental understanding of symmetry-protected topological phases with potential applications for quantum information technology~\cite{Hatsugai1993PRL, Kane2010RMP, Sone2019PRL, Hughes2011PRB, Hatsugai1993PRB, yang2020PRX}.

Non-spatial symmetries have been the subject of extensive study in both linear~\cite{PhysRevD.103.094035, PhysRevD.107.054030, rechtsman2013photonic, sarma2015majorana} and nonlinear systems~\cite{sone2020topological, zilberberg2018N, noh2018NP, mukherjee2021PRX, plotnik2014NM, noh2017NP, lumer2013PRL2, bekenstein2015NP, sharabi2018PRL, lamhot2010PRL, PhysRevResearch.4.013195, PhysRevB.108.075412, PhysRevLett.132.126601}. For instance, time-reversal and parity symmetries have been thoroughly explored in the context of mechanical, electrical and photonic structures~\cite{Tempelman2021PRB, ablowitz2022PDNP, PhysRevLett.123.053902, abdullaev2014JPCS, lumer2013PRL, ablowitz2013PRL, lu2014JPMT, ablowitz2022PRL}, enabling novel designs of microcavities~\cite{fan2012all}, circuit metamaterials~\cite{schindler2011experimental}, and plasmonic waveguides~\cite{benisty2011implementation}. Other non-spatial symmetries, such as particle-hole symmetry and chiral symmetry, have been studied in linear systems and special nonlinear systems, such as the Kerr and beyond-Kerr nonlinear interactions~\cite{ozawa2019RMP, jezequel2022PRB, Bomantara2017PRB, jezequel2022estimating} that preserve $\textbf{U}(1)$-gauge symmetry on the nonlinear wave functions~\cite{sone2023nonlinearity, Sone2022PRR}, and Rock-Paper-Scissors cycles in zero-sum games~\cite{Frey2020PRL, Gong2022PRB, Geiger2018PRE, Yoshida2021PRE, Hatsugai2022SR}. However, the rigorous definition and establishment of (anti-unitary) particle-hole symmetry and (unitary) chiral symmetry have not been addressed in nonlinear systems that do not exhibit $\textbf{U}(1)$-gauge symmetry in their nonlinear wave functions.

Nonlinear interactions are ubiquitous in nature, such as nonlinear mechanical~\cite{Tempelman2021PRB, Lo2021PRL, Liu2017RSA, Rosa2019PRL, Ma2023PRL} and electrical structures~\cite{Hohmann2023PRR, Wang2019NC}, circadian rhythms of living
cells~\cite{Edgar2012N}, and quantum fluids in optical lattices~\cite{Alyatkin2021NC}. These nonlinear mechanisms possess unique features that cannot be observed in linear systems, including soliton propagation~\cite{Parker2023PRE, PhysRevE.108.024214, pan2023PRL}, nonlinear localized modes~\cite{vakakis2001normal}, bifurcation~\cite{Fruchart2021N}, and chaos~\cite{Eckmann1985RMP}. Given the significant influence of particle-hole symmetry and chiral symmetry on linear systems, it is intriguing to ask what happens when these two symmetries encounter nonlinear dynamics.

In this work, we study the nonlinear dynamics using generalized nonlinear Schr\"{o}dinger equations. We extend the concept of particle-hole symmetry and chiral symmetry to nonlinear dynamics, where the nonlinear modes do not necessarily possess $\textbf{U}(1)$-gauge symmetry. We investigate the nontrivial consequences on nonlinear topological physics that are derived from particle-hole and chiral symmetries. Our motivation derives from the history of linear topological insulators, where although extensive research had been conducted in topological physics, there continued to be fundamental importance in the ``non-spatial" classification of symmetry-protected topological phases~\cite{RevModPhys.88.035005, schnyder2008classification}. Such a ``ten-fold" non-spatial-symmetry classification significantly enhances the potential application of topological physics, such as topological quantum computation.

To explore the impact of particle-hole and chiral symmetries on nonlinear topological physics, we investigate the geometric phase of nonlinear normal modes under the adiabatic evolution of system parameters. 
We find that, interestingly, this adiabatic geometric phase can be quantized by chiral symmetry, and defines the topologically trivial and non-trivial phases of the generalized nonlinear Schr\"{o}dinger equations. In the topologically non-trivial phase, nonlinear modes appear at the open boundaries of the system. These modes possess ``topological protection" as they show resistance against disturbances to both the modes themselves and the nonlinear interactions. Furthermore, due to chiral symmetry, the frequencies of nonlinear boundary modes are pinned at zero. Consequently, these topologically robust modes are static in time. Finally, we use a Lotka-Volterra model~\cite{RevModPhys.43.231, Bunin2017PRE, brenig1988PLA} to demonstrate the practical application of our results.




The organization of this paper is as follows. Section \uppercase\expandafter{\romannumeral2} defines the model, which is the generalized nonlinear Schr\"{o}dinger equations. Section \uppercase\expandafter{\romannumeral3} derives the adiabatic geometric phase in nonlinear normal modes. This adiabatic phase is separated into two parts: the Berry phase of nonlinear normal modes, and the component that is unique to nonlinear systems. In Section \uppercase\expandafter{\romannumeral4}, we discuss two types of non-spatial symmetries: particle-hole symmetry and chiral symmetry. Notably, we demonstrate the quantization of the Berry phase of nonlinear normal modes under chiral symmetry. Section \uppercase\expandafter{\romannumeral5} investigates the topological phases in both linear and nonlinear regimes and discusses the nonlinear topological boundary modes.

\section{The model}

\begin{figure}[htbp]
\centering
\includegraphics[scale=0.57]{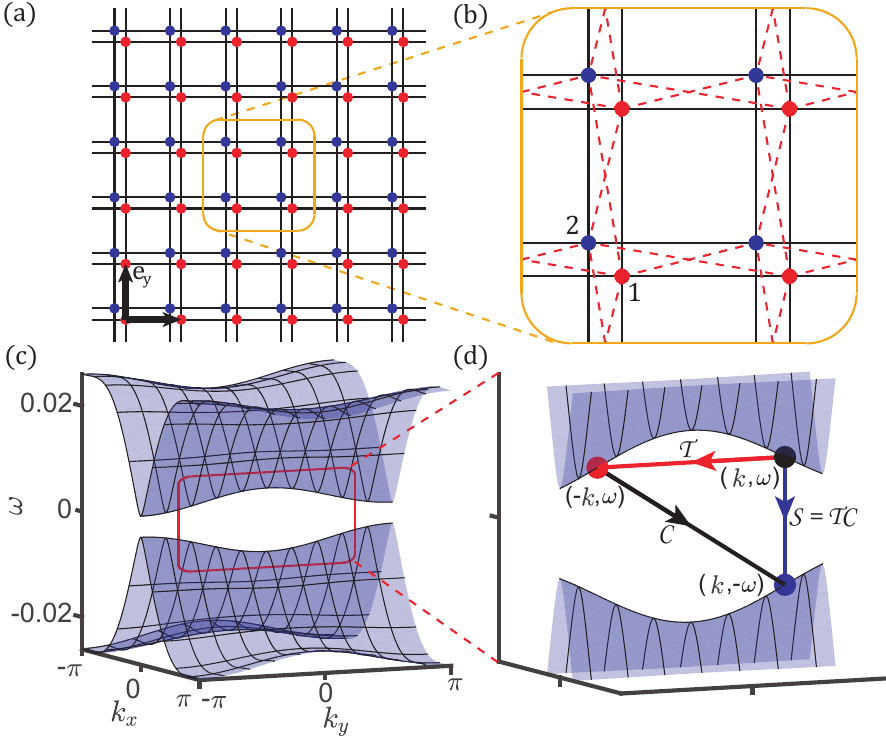}
\caption{The model of nonlinear dynamics. (a) The diatomic two-dimensional square lattice, with the primitive vectors $\hat{e}_x$ and $\hat{e}_y$ represented by the black arrows. (b) Red and blue dots mark the $1$ and $2$ sites within the unit cell. The nonlinear interactions within $1$ or $2$ sites (between $1$ and $2$ sites) are denoted by black solid lines (red dashed lines). We use the quadratic nonlinearity as the example for our nonlinear topological physics. (c) The nonlinear band structure~\cite{Fronk2017JVA, LVSI} of the model with the quadratic interactions specified in Eqs. (\ref{BdGLV}), where $\epsilon=0.001$. $g_x=0.9$, $g_y=0.1$, and mode amplitudes $A=10\epsilon$. (d) Enlarged nonlinear band structure in (c) illustrates the relationships among time-reversal, particle-hole, and chiral partner modes. 
}\label{fig1}
\end{figure}

Nonlinear dynamics, including mechanical lattices~\cite{Chaunsali2021PRB, Chaunsali2019PRB, PhysRevE.108.054224}, electrical structures~\cite{Hohmann2023PRR}, and nonlinear materials~\cite{ma2021topology, PhysRevB.108.195142}, 
can be described by generalized nonlinear Schr\"{o}dinger equations. Unlike Bose-Einstein condensates and Kerr-nonlinear optics, nonlinear interactions in these classical structures do not possess $\textbf{U}(1)$-gauge symmetry in their nonlinear normal modes~\cite{vakakis2001normal, PhysRevLett.123.053902}. To establish particle-hole symmetry and chiral symmetry in generalized nonlinear Schr\"{o}dinger equations, we consider a two-dimensional square lattice with the primitive vectors $\hat{e}_x$ and $\hat{e}_y$, as illustrated in Fig. \ref{fig1}. The unit cells are identified by two integers $n_x$ and $n_y$, corresponding to their positions $\bm{r} = n_x\hat{e}_x+n_y\hat{e}_y$. Within each unit cell, there exist two classical fields, namely $\Psi^{(1)}_{\bm{r}}$ and $\Psi^{(2)}_{\bm{r}}$. The dynamics of these classical variables are governed by the following generalized nonlinear Schr\"{o}dinger equations,  
\begin{eqnarray}\label{GNLS}
\mathrm{i}\partial_t \Psi^{(1)}_{\bm{r}} & = & +\sum_{\langle \bm{r}',\bm{r}\rangle,i'=1,2}U(\Psi^{(1)}_{\bm{r}},\Psi^{(i')}_{\bm{r}'}), \nonumber \\
\mathrm{i}\partial_t \Psi^{(2)}_{\bm{r}} & = & -\sum_{\langle \bm{r}',\bm{r}\rangle,i'=1,2}U(\Psi^{(2)}_{\bm{r}},\Psi^{(i')}_{\bm{r}'}).
\end{eqnarray}
Here, $\langle \bm{r}',\bm{r}\rangle$ denotes the nearest-neighbor sites $\bm{r}$ and $\bm{r}'$ in the square lattice. $U(\Psi^{(i)}_{\bm{r}},\Psi^{(i')}_{\bm{r}'})$ is a real-coefficient quadratic polynomial that describes the nonlinear interaction between the classical fields $\Psi^{(i)}_{\bm{r}}$ and $\Psi^{(i')}_{\bm{r}'}$:
\begin{eqnarray}\label{BdGLV}
 & {} & U(\Psi^{(i)}_{\bm{r}},\Psi^{(i')}_{\bm{r}'}) = (\epsilon+\Psi^{(i)}_{\bm{r}}) \Psi^{(i')}_{\bm{r}'} \bigg[ (n_x-n_x') (1-\delta_{ii'}) \nonumber \\
 & {} & 
+\bigg(-\delta_{n_x,n_x'}\delta_{n_y,n_y'}+\sum_{j=x,y}g_j |n_j-n_j'| \bigg)\delta_{ii'}\bigg],
\end{eqnarray}
where $\epsilon$ represents the linear on-site potential, and $g_j$ for $j=x,y$ accounts for the nonlinear effects between the nearest-neighbor classical fields. These constant parameters are positive and real numbers, with $\epsilon\ll 1$ utilized to emphasize the significance of nonlinear effects. This nonlinear interaction is pictorially represented by the black solid and red dashed lines in Fig. \ref{fig1}(b). 

This model is capable of describing a number of physical systems. When the amplitude of the classical fields is much smaller than $\epsilon$, this classical model is in the linear regime. The topological boundary modes in this linearized classical model can mimic the quantum fermionic edge states within the Bogoliubov-de Gennes Hamiltonian of topological superconductors~\cite{Sato2017RPP, Ryu2002PRL}. When the mode amplitude becomes comparable or greater than $\epsilon$, the model is highly nonlinear in the classical field variables. In this regime, the \emph{static solutions} of this nonlinear Schr\"{o}dinger-type dynamics can mimic the \emph{static solutions} of the Lotka-Volterra model~\cite{RevModPhys.43.231, Adachi2022CP, Ludwig1978JAE}, which we elaborate later in Section \uppercase\expandafter{\romannumeral5}. 

This nonlinear interaction, as represented by Eq. (\ref{BdGLV}), yields the inequality:
\begin{eqnarray}\label{U1}
U(e^{\mathrm{i}\theta}\Psi^{(i)}_{\bm{r}},e^{\mathrm{i}\theta}\Psi^{(i')}_{\bm{r}'})\neq e^{\mathrm{i}\theta}U(\Psi^{(i)}_{\bm{r}},\Psi^{(i')}_{\bm{r}'}),
\end{eqnarray}
indicating that the $\textbf{U}(1)$-gauge symmetry in the nonlinear wave functions is broken. 
To further clarify this statement, 
we denote the nonlinear wave function as the following form
\begin{eqnarray}
\Psi = (\ldots,\Psi^{(1)}_{\bm{r}}, \Psi^{(2)}_{\bm{r}},\ldots)^\top.
\end{eqnarray}
This is a column vector with $2N$ components, where $N$ is the number of diatomic unit cells, and $\top$ denotes matrix transpose. In a concise representation, Eqs. (\ref{GNLS}) can be written as 
\begin{eqnarray}
\mathrm{i}\partial_t\Psi = H(\Psi)
\end{eqnarray}
where $H(\Psi)$ is a nonlinear mapping of $\Psi$, and is called the nonlinear Hamiltonian. It is a $2N\times 1$ column vector with each element given by Eq. (\ref{BdGLV}). Again, we emphasize that the nonlinear Hamiltonian does not preserve the $\textbf{U}(1)$-gauge symmetry in the nonlinear modes by showing  
\begin{eqnarray}\label{U2}
H(e^{\mathrm{i\theta}}\Psi)\neq e^{\mathrm{i}\theta}H(\Psi).
\end{eqnarray}


In Section \uppercase\expandafter{\romannumeral4}, we will demonstrate that the nonlinear model in Eq. (\ref{GNLS}) respects particle-hole symmetry and chiral symmetry.

\section{Adiabatic geometric phase of nonlinear normal modes}

In this section, we derive the geometric phase that arises from the adiabatic evolution of nonlinear normal modes in the model presented in Eqs. (\ref{GNLS}). The adiabatic geometric phase is especially useful in characterizing the topological phases and predicting the existence of topological boundary modes in nonlinear dynamics.

When the amplitudes of classical fields are much smaller than $\epsilon$, the nonlinear Hamiltonian can be linearized into a two-band matrix Hamiltonian $\hat{H}_{\bm{k}}$~\cite{Ryu2002PRL}. This Hamiltonian describes the oscillation of classical variables governed by the eigenfrequency $\omega=\omega(\bm{k})$, where the wavevector $\bm{k}=(k_x,k_y)$ resides in the two-dimensional Brillouin zone.

Nonlinearities become increasingly important as mode amplitude rises, making nonlinear normal modes significantly deviate from sinusoidal waves of linear systems~\cite{Hohmann2023PRR, Tang2023FOP, Tsapalis2016PRD, vakakis2001normal}. Specifically, nonlinear normal modes with the plane-wave format satisfy the expression 
\begin{eqnarray}\label{NNM}
\Psi_{\bm{k},\omega} = (\Psi^{(1)}_{\bm{k}}(\bm{k}\cdot \bm{r}-\omega t),\Psi^{(2)}_{\bm{k}}(\bm{k}\cdot \bm{r}-\omega t+\phi_{\bm{k}}))^\top
\end{eqnarray}
according to the nonlinear extension of the Bloch theorem~\cite{Zhou2022NC, vakakis2001normal}. Here, $\Psi^{(1)}_{\bm{k}}(\theta)$ and $\Psi^{(2)}_{\bm{k}}(\theta)$ represent the $2\pi$-periodic non-sinusoidal functions, while $\phi_{\bm{k}}$ characterizes the relative phase between these wave components. The frequencies of these nonlinear normal modes, denoted by $\omega = \omega(\bm{k}, A)$, are influenced by both the wavevectors $\bm{k}$ and mode amplitudes $A$, deviating from their linear counterparts. It is worth emphasizing that in nonlinear systems, the number of nonlinear modes can exceed the degrees of freedom, allowing nonlinear localized modes~\cite{chong2021nonlinear, vakakis2001normal} and ``looped band structures"~\cite{chen2011many, PhysRevB.102.115411, PhysRevB.96.121406, koller2016nonlinear} to emerge from the effect of bifurcation. However, in this work, our scope is limited to the simple case that bifurcation does not occur, and these additional excitations do not emerge from the nonlinear dynamics. Thus, plane-wave nonlinear normal modes can be uniquely defined based on their amplitude, wavevector, and frequency, allowing the nonlinear system to be effectively described as a ``two-band nonlinear model."

We adiabatically evolve the plane-wave nonlinear normal mode as the wavevector $\bm{k} = \bm{k}(t)$ follows a closed trajectory $C$ in the Brillouin zone~\cite{RevModPhys.82.1959}. Based on the nonlinear extension of the adiabatic theorem~\cite{PhysRevLett.90.170404, Pu2007PRL}, at time $t$, the nonlinear normal mode follows the ansatz 
\begin{eqnarray}
\Psi=\Psi_{\bm{k}(t),\omega}\left(-\int_0^t\omega(t',\bm{k}(t'))dt'-\gamma(t)\right),
\end{eqnarray}
where $\gamma(t)$ denotes the phase shift of the nonlinear normal mode during the adiabatic evolution. 
When the wavevector $\bm{k}$ completes a closed-loop travel in the Brillouin zone, the wave function acquires a phase $\gamma_C$, known as the adiabatic geometric phase~\cite{PhysRevA.81.052112, litvinets2006JPMG}.

As per the Whitham modulation theory~\cite{abeya2023JPMT, minzoni2016PDNP, biondini2023whitham}, during the adiabatic evolution of the wavevector $\bm{k}$ in reciprocal space, the mode amplitude $A$ and relative phase $\phi_{\bm{k}}$ (as defined in Eq. (\ref{NNM})) should change slowly. Consequently, the adiabatic phase, denoted by $\gamma_C$, can be separated into two components: $\gamma_C = \gamma_C^{(\rm B)}+\gamma_C^{(\rm NL)}$. The first term, $\gamma_C^{(\rm B)}$, is referred to as the Berry phase of nonlinear normal modes, which originates from the change in the relative phase $\phi_{\bm{k}}$. On the other hand, the second term, $\gamma_C^{(\rm NL)}$, arises from the change in the mode amplitude. In the upcoming discussion, we will examine $\gamma_C^{(\rm B)}$ and $\gamma_C^{(\rm NL)}$ individually.

As described in Appendix A, the derivation of the Berry phase of nonlinear normal modes involves computing the evolution of the relative phase $\phi_{\bm{k}}$ as the wavevector $\bm{k}$ undergoes adiabatic changes:
\begin{eqnarray}\label{gamma1}
\gamma_C^{(\rm B)} = \oint_{C}\frac{\sum\limits_{l}\bigg(l|\psi^{(2)}_{l\bm{k}}|^2 \nabla_{\bm{k}}\phi_{\bm{k}}-\mathrm{i}\sum\limits_{i=1,2}\psi^{(i)*}_{l\bm{k}}\nabla_{\bm{k}}\psi^{(i)}_{l\bm{k}}\bigg)}{\sum\limits_{l'}l' \left(|\psi^{(1)}_{l'\bm{k}}|^2+|\psi^{(2)}_{l'\bm{k}}|^2\right)}\cdot d\bm{k}.\nonumber \\
\end{eqnarray}
Here, $\psi^{(i)}_{l\bm{k}}= (2\pi)^{-1}\int_0^{2\pi} e^{-\mathrm{i} l\theta} \Psi^{(i)}_{\bm{k}} d\theta$ is the 
$l$-th Fourier component of the nonlinear wave, $\Psi^{(i)}_{\bm{k}}$, with $i=1,2$. We emphasize that for Schr\"{o}dinger equations with linear~\cite{Niu2010RMP} or nonlinear interactions that exhibit $\textbf{U}(1)$-gauge symmetry~\cite{tuloup2020PRB} in their eigenstates, the eigenmodes can be described using fundamental harmonics only. This reduces Eq. (\ref{gamma1}) to the Berry phase in Ref.~\cite{PhysRevA.81.052112}, $\gamma^{(\rm B)}_{C, {\rm linear}} = \mathrm{i}\oint_C d\bm{k}\cdot \langle \Psi_{\bm{k},\omega}|\nabla_{\bm{k}}|\Psi_{\bm{k},\omega}\rangle$ (see Appendix A for details). Usually, $\gamma_C^{(\rm B)}$ is \emph{not quantized} in systems without symmetry constraints. However, $\gamma_C^{(\rm B)}$ can be quantized 
if non-spatial symmetries are incorporated in the system, which we address in Section \uppercase\expandafter{\romannumeral4}(D).

The Whitham modulation theory~\cite{abeya2023JPMT, minzoni2016PDNP, biondini2023whitham} also indicates that mode amplitude should change during adiabatic evolution. This effect gives rise to an additional contribution to the adiabatic geometric phase~\cite{PhysRevA.81.052112}, denoted as $\gamma_C^{(\rm NL)}$. However, as we will demonstrate in Section \uppercase\expandafter{\romannumeral4}(C), chiral symmetry imposes a constraint that causes the amplitudes of the two wave components of a nonlinear normal mode to be equal. This constraint results in the mode amplitude staying unchanged, up to the normalization factor of the wave function. Consequently, we have found that the adiabatic geometric phase $\gamma_C^{(\rm NL)}$ vanishes under such constraints. This vanishing result is derived in Ref.~\cite{PhysRevA.81.052112} and in Appendix A. As a result, the Berry phase of nonlinear normal modes, $\gamma_C^{(\rm B)}$, is the only contribution to the adiabatic geometric phase. Under the constraint of chiral symmetry, this Berry phase takes quantized values and classifies the topological phases of the nonlinear dynamics. 

\section{Particle-hole symmetry, chiral symmetry, and topological index of nonlinear systems}

In this section, we establish particle-hole symmetry and chiral symmetry for nonlinear dynamics. We show that chiral symmetry can be used to quantize the Berry phase of nonlinear normal modes in the model described by Eqs. (\ref{GNLS}). This quantized index serves as a topological invariant and defines the topologically trivial and non-trivial phases of the nonlinear dynamics. 

We briefly review time-reversal symmetry that has been well-established in nonlinear dynamics~\cite{RevModPhys.88.035002, vakakis2001normal, ozawa2019RMP, christodoulides2018parity}. The nonlinear system is time-reversal invariant if the equations of motion remain unchanged under the time-reversal transformation $(\Psi_{\bm{r}}^{(1)}(t),\Psi_{\bm{r}}^{(2)}(t))\to (\Psi_{\bm{r}}^{(1)*}(-t),\Psi_{\bm{r}}^{(2)*}(-t))$. This invariance arises from the real-coefficient polynomials of the nonlinear interactions in terms of the field variables. The formal expression of time-reversal symmetry is $\mathcal{T}H(\Psi)- H(\mathcal{T}\Psi)=0$, where $H(\Psi)$ is the nonlinear Hamiltonian, and the time-reversal operator, $\mathcal{T}=\mathcal{K};t\to-t$, involving complex conjugation and reversing the sign of time. Here, $H(\mathcal{T}\Psi)$ means that we perform the time-reversal operation on the nonlinear wave $\Psi$, and then we operate on $\mathcal{T}\Psi$ using the nonlinear Hamiltonian (nonlinear mapping) $H$. This symmetry implies that for a nonlinear wave $\Psi_{\bm{k},\omega}$ in a time-reversal symmetric model, there exists a time-reversed partner solution with the wavevector $-\bm{k}$, denoted as $\Psi_{-\bm{k},\omega}=\mathcal{T}\Psi_{\bm{k},\omega} = (\Psi^{(1)*}_{\bm{k}}(\bm{k}\cdot \bm{r}+\omega t),\Psi^{(2)*}_{\bm{k}}(\bm{k}\cdot \bm{r}+\omega t+\phi_{\bm{k}}))^\top$, as pictorially indicated by the red arrow and dot in Fig. \ref{fig1}(d).

The time-reversal operator satisfies $\mathcal{T}^2=1$ and is anti-unitary, aligning with the operator for linear Schr\"{o}dinger equation of spinless particles~\cite{Haim2019PR, Zhang2011RMP}. In the linear regime, time-reversal symmetry simplifies to the conventional form, $\mathcal{T}\hat{H}_{\bm{k}}\mathcal{T}^{-1} = \hat{H}_{-\bm{k}}$, where $\hat{H}_{\bm{k}}$ represents the linearized Hamiltonian in reciprocal space.

\subsection{Particle-hole symmetry of nonlinear dynamics}

Here, we introduce the particle-hole operator and its corresponding symmetry for nonlinear systems. 

In a particle-hole symmetric system, the motion equations in Eqs. (\ref{GNLS}) remain unchanged under the particle-hole transformation, $(\Psi_{\bm{r}}^{(1)},\Psi_{\bm{r}}^{(2)})\to (\Psi_{\bm{r}}^{(2)*},\Psi_{\bm{r}}^{(1)*})$. This invariance is equivalently captured by the constraint,
\begin{eqnarray}\label{C}
\mathcal{C} H(\Psi) + H(\mathcal{C} \Psi)=0, \qquad
\mathcal{C} =
{\rm I}_N\otimes\sigma_x\mathcal{K},
\end{eqnarray}
where $\mathcal{C}$ is the particle-hole operator, ${\rm I}_N$ is the $N\times N$ identity matrix, and $\sigma_x$ is the Pauli matrix. One important characteristic of the particle-hole operator we have defined is its \emph{anti-unitary} property, which arises from the complex conjugation involved in the operator. This property, in turn, results in the reversal of the frequency and momentum of the original nonlinear wave (see Eq. (\ref{NNM})) in the system under particle-hole transformation. Therefore, for a nonlinear normal mode $\Psi_{\bm{k},\omega}$ with the wavevector $\bm{k}$ and frequency $\omega$, the particle-hole-symmetric model has a corresponding nonlinear sister solution with a wavevector $-\bm{k}$ and frequency $-\omega$, denoted as 
\begin{eqnarray}\label{NNM2}
 & {} & \Psi_{-\bm{k},-\omega}=\mathcal{C}\Psi_{\bm{k}, \omega} =\nonumber \\
 & {} & (\Psi^{(2)*}_{\bm{k}}(\bm{k}\cdot\bm{r}-\omega t),\Psi^{(1)*}_{\bm{k}}(\bm{k}\cdot\bm{r}-\omega t-\phi_{\bm{k}}))^\top,
\end{eqnarray}
as depicted by the black arrow in Fig. \ref{fig1}(d). The particle-hole operator satisfies $\mathcal{C}^2=1$ and is anti-unitary, aligning with the operator defined for linear systems~\cite{Ryu2002PRL, Zhang2011RMP}. 

These results, including the particle-hole symmetry presented in Eq. (\ref{C}) and the particle-hole-partner mode in Eq. (\ref{NNM2}), naturally apply to linear and nonlinear systems that respect the $\textbf{U}(1)$-gauge symmetry. Furthermore, in these $\textbf{U}(1)$-symmetric systems, Eqs. (\ref{C}) can be reduced to the well-studied format defined in reciprocal space, namely $\mathcal{C}\hat{H}_{\bm{k}}\mathcal{C}^{-1} = -\hat{H}_{-\bm{k}}$.

\subsection{Chiral symmetry of nonlinear dynamics}

Chiral symmetry naturally arises in nonlinear systems when both time-reversal and particle-hole symmetries are present. The model described by Eqs. (\ref{GNLS}) remains unchanged under the chiral transformation $(\Psi_{\bm{r}}^{(1)}(t),\Psi_{\bm{r}}^{(2)}(t))\to (\Psi_{\bm{r}}^{(2)}(-t),\Psi_{\bm{r}}^{(1)}(-t))$. Mathematically, chiral symmetry is expressed as the constraint on the nonlinear Hamiltonian given by 
\begin{eqnarray}\label{S}
\mathcal{S}H(\Psi) + H(\mathcal{S} \Psi)=0, \qquad \mathcal{S}=\mathcal{T}\cdot\mathcal{C},
\end{eqnarray}
where $\mathcal{S}=\mathcal{T}\cdot\mathcal{C}$ represents the chiral symmetry operator that combines the effects of time-reversal and particle-hole transformations. It is worth noting that the resulting chiral symmetry operator, which is a combination of time-reversal and particle-hole symmetry operators expressed as $\mathcal{S}=\mathcal{T}\cdot\mathcal{C}$, is \emph{unitary} in nature. This is due to the anti-unitary nature of both time-reversal and particle-hole symmetry operators, which cancels out to produce a unitary chiral symmetry operator that is critical for characterizing the topological properties of the nonlinear system, which we address in the following subsection. Thus, for a nonlinear normal mode $\Psi_{\bm{k},\omega}$ with frequency $\omega$, chiral symmetry predicts the existence of a partner solution with frequency $-\omega$, denoted as 
\begin{eqnarray}\label{NNM3}
 & {} & \Psi_{\bm{k},-\omega}=\mathcal{S}\Psi_{\bm{k},\omega}=\nonumber \\
 & {} & (\Psi^{(2)}_{\bm{k}}(\bm{k}\cdot \bm{r}+\omega t), \Psi^{(1)}_{\bm{k}}(\bm{k}\cdot \bm{r}+\omega t-\phi_{\bm{k}}))^\top.
\end{eqnarray}
This relationship is depicted by the green arrow and dot in Fig. \ref{fig1}(d). This chiral operator is unitary and yields $\mathcal{S}^2=1$, agreeing perfectly with the chiral operator for linear topological insulators.

These results, including the chiral symmetry presented in Eq. (\ref{S}) and the chiral-partner mode in Eq. (\ref{NNM3}), naturally apply for linear and nonlinear systems that respect the $\textbf{U}(1)$-gauge symmetry. Furthermore, in these $\textbf{U}(1)$-symmetric systems, Eqs. (\ref{S}) can be reduced to the well-studied format~\cite{Haim2019PR, Ryu2010NJP, Dumitrescu2014PRB} defined in reciprocal space, namely $\mathcal{S}\hat{H}_{\bm{k}}\mathcal{S}^{-1}=-\hat{H}_{\bm{k}}$.
Finally, our study highlights an important aspect of chiral-symmetric nonlinear systems. We find that the chiral-symmetric partner nonlinear mode swaps the two wave components of a nonlinear normal mode. This swapping leads to a result where the two wave components in a nonlinear normal mode share the same mode amplitude.

Our nonlinear model exhibits time-reversal, particle-hole, and chiral symmetries, which is the extension of symmetry class BDI defined in the ten-fold classification of linear topological insulators~\cite{Ryu2010NJP}.

\subsection{Berry phase of nonlinear normal modes quantized by chiral symmetry}

In a purely linear Schr\"{o}dinger equation, the summation of Berry phases across all energy bands is always zero due to the topological triviality of the fiber bundle~\cite{RevModPhys.82.1959} associated with a complete and orthogonal set of eigenbasis in a matrix Hamiltonian~\cite{RevModPhys.82.1959}. 
Combined with chiral symmetry, this sum rule naturally results in the quantization of linear Berry phase~\cite{Zhou2020PRR}. However, this conclusion does not hold for nonlinear dynamics, because nonlinear normal modes do not necessarily have completeness and orthogonality when matrix analysis fails. Therefore, the quantization of Berry phase of nonlinear normal modes in a chiral-symmetric nonlinear system, as expressed in Eq. (\ref{gamma1}), remains an open question.

Here, we demonstrate that under chiral symmetry, the Berry phase of nonlinear normal modes still remains quantized for nonlinear systems. Moreover, the quantization of the Berry phase of nonlinear normal modes suggests that it can serve as a potential topological index for characterizing the topological phases of the underlying nonlinear dynamics. 

To demonstrate that chiral symmetry is still capable of quantizing the Berry phase of nonlinear normal modes, we consider the nonlinear normal mode $\Psi_{\bm{k},-\omega} = \mathcal{S}\Psi_{\bm{k},\omega}$, which is the chiral-partner mode of $\Psi_{\bm{k},\omega}$. We perform an adiabatic evolution on this chiral-partner mode by slowly varying the wavevector $\bm{k}(t)$ along a closed-loop trajectory $C$ in the Brillouin zone. When we adiabatically evolve the chiral-symmetric partner mode, $\Psi_{\bm{k},-\omega}$, this mode can be considered as having a frequency of $\omega$ but with the arrow of time reversed. Consequently, the mode acquires a term $\gamma(t)$ in its phase variable, via
\begin{eqnarray}
\Psi(t) = \Psi_{\bm{k}(t), -\omega}\left(\int_0^t\omega(t',\bm{k}(t'))dt'+\gamma(t)\right).
\end{eqnarray}
Substituting this result into the nonlinear motion equations, namely $\mathrm{i}\partial_t\Psi_{\bm{k}(t),-\omega}=H(\Psi_{\bm{k}(t),-\omega})$, we can compute the geometric phase when the wavevector $\bm{k}$ travels along the trajectory $C$. This adiabatic evolution allows the chiral-symmetric nonlinear mode to obtain the adiabatic geometric phase, $\gamma_C = \gamma_C^{(\rm B)}$ (note that $\gamma_C^{(\rm NL)}=0$ in chiral-symmetric systems, as shown in Appendix A), from which the Berry phase of nonlinear normal modes, $\gamma_C^{(\rm B)}$, is obtained:
\begin{eqnarray}\label{gamma2}
\gamma_C^{(\rm B)} = \oint_{C}\frac{\sum\limits_{l}\bigg(l|\psi^{(1)}_{l\bm{k}}|^2 \nabla_{\bm{k}}\phi_{\bm{k}}+\mathrm{i}\sum\limits_{i=1,2}\psi^{(i)*}_{l\bm{k}}\nabla_{\bm{k}}\psi^{(i)}_{l\bm{k}}\bigg)}{\sum\limits_{l'}l' \left(|\psi^{(1)}_{l'\bm{k}}|^2+|\psi^{(2)}_{l'\bm{k}}|^2\right)}\cdot d\bm{k}.\nonumber \\
\end{eqnarray}
Since both Eq. (\ref{gamma1}) and (\ref{gamma2}) describe the same Berry phase of nonlinear normal modes under the same evolution trajectory, we equate them, and obtain the result 
\begin{eqnarray}\label{gammaC}
\gamma_C^{(\rm B)} = \frac{1}{2}\oint_C \nabla_{\bm{k}}\,\phi_{\bm{k}}\cdot d\bm{k} = n\pi, \quad n=0,1.
\end{eqnarray}
This equation demonstrates the quantization of the Berry phase of nonlinear normal modes under the constraint of chiral symmetry, which serves as the topological invariant of the considered nonlinear dynamics. $n=0,1$ correspond to the nonlinear topologically trivial and non-trivial phases, respectively. Due to the quantized nature of this Berry phase, it cannot change continuously upon the variations of system parameters, including the mode amplitudes and coupling parameters. Below, we leverage the invariance of this topological number to investigate the corresponding nonlinear topological edge modes.

\section{Nonlinear topological phases and boundary modes}

We have demonstrated the topological invariance of the Berry phase of nonlinear normal modes using chiral symmetry. In this section, we exemplify the impact of this topological number on the nonlinear physics, where the interactions are specified as the example in Eqs. (\ref{BdGLV}) with the quadratic nonlinearities. Specifically, we investigate the topological phases and the corresponding boundary physics of the system in both the linear and nonlinear regimes.

\begin{figure}[htbp]
\centering
\includegraphics[scale=0.58]{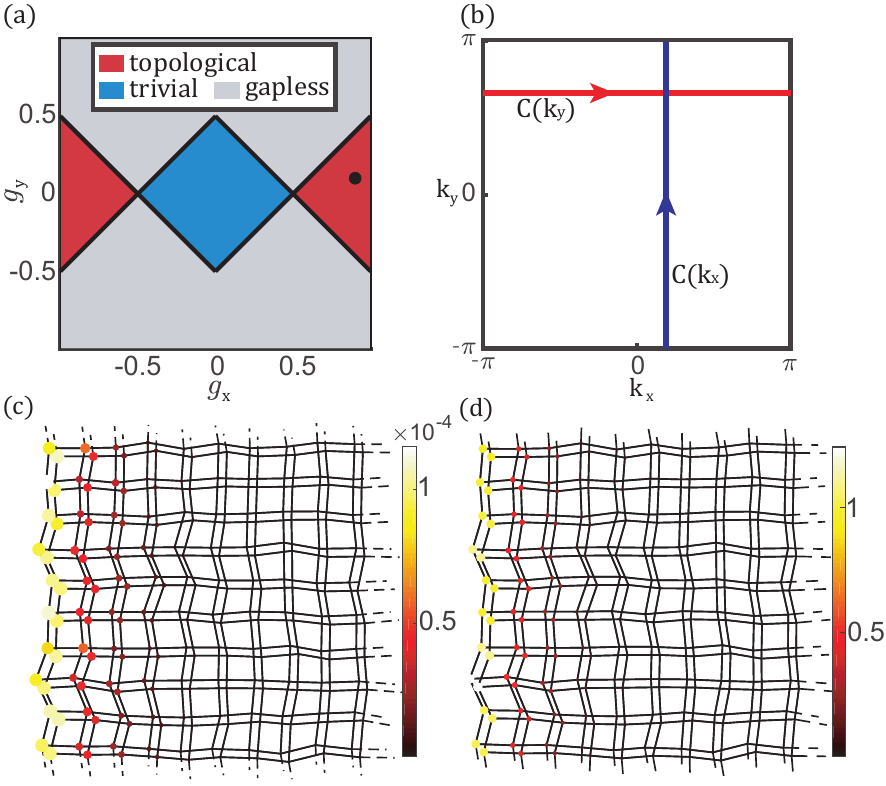}
\caption{Topological phases and boundary modes. (a) Nonlinear topological phase diagram of the model with quadratic nonlinearities. (b) Brillouin zone with the integration trajectories of Berry phases of nonlinear normal modes indicated. (c) Disks' size and color represent the spatial profile of the weakly nonlinear topological edge mode. (d) Spatial distribution of the topological boundary mode in the highly nonlinear regime.
}\label{fig2}
\end{figure}

\subsection{Topological phases and boundary modes in the linear regime}

When the quadratic nonlinearities are small compared to the linear parts, the nonlinear system can be approximated as linear Schr\"{o}dinger equations. This regime is valid when mode amplitudes $A\ll\epsilon$. This allows us to perform a momentum-space decomposition and simplify the linearized equations of motion using the equation, $\mathrm{i}\partial_t\Psi_{\bm{k},\omega}=\hat{H}_{\bm{k},\omega}\Psi_{\bm{k},\omega}$, where the the linear wave function is represented as $\Psi_{\bm{k},\omega}=(\Psi^{(1)}_{\bm{k}}, \Psi^{(2)}_{\bm{k}})^\top$, the matrix Hamiltonian, $\hat{H}_{\bm{k},\omega}$, reads
\begin{eqnarray}\label{C13}
\hat{H}_{\bm{k},\omega} = 2\epsilon \sigma_y \sin k_x + \epsilon\sigma_z\bigg(-1+2\sum_{j=x,y}g_j\cos k_j\bigg),\quad
\end{eqnarray}
where $\bm{k}=(k_x, k_y)$ is the wavevector within the two-dimensional Brillouin zone, and $\sigma_{x,y,z}$ are Pauli matrices. This Hamiltonian possesses chiral symmetry, namely a unitary chiral symmetry operator $\mathcal{S}=\sigma_x$ anti-commutes with it~\cite{Ryu2002PRL, Haim2019PR}: $\{\sigma_x,\hat{H}_{\bm{k},\omega}\}=0$.

The Hamiltonian $\hat{H}_{\bm{k},\omega}$ has multiple phases. Specifically, when the interaction parameters are in the grey region of Fig. \ref{fig2}(a) (i.e., the interaction parameters satisfy $|g_y|\ge |g_x-1/2|\bigcup|g_y|\ge |g_x+1/2|$), the system becomes gapless, as the linear band structures touch at a pair of zero-frequency points. As a result, the linear Berry phase becomes singular and ill-defined when the integration trajectory passes through these gapless points. When the linear bands touch at the gapless points, the corresponding static eigenstates are given by 
\begin{eqnarray}\label{Weyl}
\Psi_{r}(t) = (1, \pm1)^\top e^{\mathrm{i}\bm{k}_w\cdot\bm{r}}/\sqrt{2},
\end{eqnarray}
whose frequency and wavevector are $\omega=0$ and $\pm\bm{k}_w$, respectively. For the parameters that lie in the region of $|g_y|\ge |g_x-1/2|$, we have the wavevectors $\bm{k}_w = (0,\arccos[(1-2g_x)/2g_y])$. When the parameters are in the region $|g_y|\ge |g_x+1/2|$, the wavevectors are given by $\bm{k}_w = (\pi,\arccos[(1+2g_x)/2g_y])$. We note that the zero-frequency nature of these gapless bulk states stems from the chiral symmetry~\cite{Sticlet2012Thesis}.

On the other hand, in the blue region of Fig. \ref{fig2}(a) (the parameters yield $|g_y|<|g_x-1/2| \bigcap|g_y|<|g_y+1/2|$), and in the red region of Fig. \ref{fig2}(a) (with the parameters that yield $-g_x+1/2<g_y<g_x-1/2 \bigcup g_x+1/2<g_y<-g_x-1/2$), the lattice is fully gapped and the linear Berry phase becomes well-defined: 
\begin{eqnarray}
\lim_{A\to 0}\gamma_C^{(\rm B)}(A) = \mathrm{i}\oint_C d\bm{k}\cdot\langle\Psi_{\bm{k},\omega}|\nabla_{\bm{k}}|\Psi_{\bm{k},\omega}\rangle, 
\end{eqnarray} 
where $C$ is the closed-loop trajectory for the adiabatic evolution of the wavevector $\bm{k}$ in the two-dimensional Brillouin zone. 
Specifically, we define two types of closed trajectories, $C(k_y): k_x=-\pi\to \pi; k_y$ and $C(k_x): k_y = -\pi \to \pi; k_x$, which integrate over all $k_x$ for a given $k_y$ and all $k_y$ for a given $k_x$, respectively. The linear Berry phase corresponding to the horizontal trajectory $C(k_y)$ is denoted as $\gamma_{C(k_y)}$, while that corresponding to the vertical trajectory $C(k_x)$ is denoted as $\gamma_{C(k_x)}$.


Due to the chiral symmetric nature of the linearized Hamiltonian~\cite{Ryu2002PRL}, these Berry phases, namely $\gamma_{C(k_y)}$ and $\gamma_{C(k_x)}$, are guaranteed to have integer multiples of $\pi$. Moreover, due to the fully gapped nature of the linear band structure, the Berry phases $\gamma_{C(k_y)}$ and $\gamma_{C(k_x)}$ remain unchanged for any $k_x$ and $k_y$ ranging from $-\pi$ to $\pi$. 
To characterize the topological phases of the linearized model, we define the ``2D polarization~\cite{Li2023PRB}" as a measure of the linear Berry phases averaged by the Brillouin zone, 
\begin{eqnarray}\label{polarization1}
\lim_{A\to 0}\bm{R}_{\rm T}(A) = \frac{1}{2\pi^2}\int_{-\pi}^{\pi} & {} & \bigg[\lim_{A\to 0}\gamma^{(\rm B)}_{C(k_y)}(A) dk_y \hat{e}_x+\nonumber \\
 & {} & \lim_{A\to 0}\gamma^{(\rm B)}_{C(k_x)}(A) dk_x \hat{e}_y\bigg].
\end{eqnarray}
When the parameters are in the red region of Fig. \ref{fig2}(a), the topological polarization is given by $\lim_{A\to 0}\bm{R}_{\rm T}(A) = \hat{e}_x$, indicating that the linear system is in the topological phase. On the other hand, when the parameters are within the blue region of Fig. \ref{fig2}(a), the topological polarization becomes $\lim_{A\to 0}\bm{R}_{\rm T}(A)=0$, indicating that the system is in the topologically trivial regime.

According to the principle of bulk-boundary correspondence in topological band theory, the behavior of the system at the boundary is determined by the topological invariant derived from the bulk bands of the lattice. In the red region of Fig. \ref{fig2}(a), the topological polarization vector takes the value $\lim_{A\to 0}\bm{R}_{\rm T}(A)=\hat{e}_x$, indicating the emergence of topologically protected boundary modes in the system. These boundary modes can be analytically solved using the Jackiw-Rebbi-type solution, 
\begin{eqnarray}\label{edgeMode1}
 & {} & \left(
 \begin{array}{c}
 \Psi^{(1)}_{\bm{r}}(t) \\ 
 \Psi^{(2)}_{\bm{r}}(t) \\
 \end{array}\right)
 =\nonumber \\
 & {} & A e^{\mathrm{i}k_y n_y}(e^{-\kappa_+(k_y)n_x}-e^{-\kappa_-(k_y)n_x})\left(
 \begin{array}{c}
1 \\ 
-1 \\
 \end{array}\right),
\end{eqnarray}
where $A$ is the mode amplitude, $k_y$ is the wave number in the transverse $y$-direction, and the spatial decay rates $\kappa_\pm(k_y)$ yield $\kappa_{\pm}(k_y)>0$. This edge mode is exponentially localized on the left open boundary of the square lattice. In contrast, in the topologically trivial phase with the parameters that lie in the blue region of Fig. \ref{fig2}(a), the topological polarization vector takes the trivial value of $\lim_{A\to 0}\bm{R}_{\rm T}(A)=0$, indicating the absence of topological boundary modes in this linear non-topological phase.

Chiral symmetry of the linearized Hamiltonian leads to an important property in the frequency spectrum of the bulk modes. This ensures that the frequencies of the bulk modes arise in pairs of $\pm\omega$, reflecting the chiral symmetry of the Hamiltonian. This property in the bulk mode frequencies has significant implications for the frequencies of topological edge states in the system. In particular, the frequency of the topological edge state is constrained to be pinned at zero. This is because, if the frequency of the topological state, $\omega$, is nonzero, then a partner topological state with the frequency of $-\omega$ must also arise at the same boundary. These two boundary states can couple and open a band gap on the lattice boundary, violating the topological protection of the edge states. Therefore, topological edge states must have zero frequency and do not evolve in time, satisfying the static condition $\partial_t \Psi^{(i)}_{\bm{r}} =0$ for $i=1,2$ and for all $\bm{r}$. The static nature of topological boundary modes remains valid for the nonlinear system as chiral symmetry extends to the fully nonlinear regime.

\subsection{Topological phases and boundary modes in the nonlinear regime}

We now consider the topological phases and the corresponding boundary modes in the Schr\"{o}dinger-type equations in the nonlinear regime.

Intriguingly, even in the strongly nonlinear regime, the system remains gapless when the parameters are in the grey-shaded regime of Fig. \ref{fig2}(a). This is because nonlinear zero-frequency bulk modes can still arise in the nonlinear regime of the system, maintaining the system's gapless nature. The nonlinear static bulk modes can be analytically obtained by imposing the condition $\partial_t\Psi_{\bm{r}}^{(i)}=0$ for $i=1,2$ and for all $\bm{r}$, because chiral symmetry assures the frequency of the nonlinear gapless mode to stay at zero. The nonlinear mode in the lattice system is described by Eq. (\ref{Weyl}), which is identical to that obtained from the linearized model. The presence of zero-frequency nonlinear bulk modes in the parameter region defined by the grey regime of Fig. \ref{fig2}(a) yields the closure of the nonlinear band gap, which we define as the ``nonlinear gapless phase" of the system.

The system with parameters in the blue and red regimes of Fig. \ref{fig2}(a) is in the nonlinear fully gapped phase, because these phases are devoid of the previously discussed zero-frequency nonlinear bulk modes. This property allows the topological numbers of the system to be well-defined and invariant as mode amplitudes rise. Thus, we define the 2D topological polarization of the lattice in the fully nonlinear regime:
\begin{eqnarray}\label{polarization2}
\bm{R}_{\rm T}(A) = \frac{1}{2\pi^2}\int_{-\pi}^{\pi}\left[\gamma^{(\rm B)}_{C(k_y)}(A) dk_y \hat{e}_x+ \gamma^{(\rm B)}_{C(k_x)}(A) dk_x \hat{e}_y\right].\nonumber \\
\end{eqnarray}
The red-shaded parameter regime shown in Fig. \ref{fig2}(a) corresponds to the nonlinear topological phase of the system described by the nonlinear Schr\"{o}dinger equation in Eqs. (\ref{GNLS}). In this regime, the topological polarization takes the form $\bm{R}_{\rm T}(A) = \hat{e}_x$, indicating the presence of nonlinear topological boundary modes at the edges of the lattice. We can calculate the spatial profile of these boundary modes by imposing static boundary conditions and solving for the corresponding eigenstates. The resulting static nonlinear mode is described by Eq. (\ref{edgeMode1}) and is exponentially localized on the left open boundary of the lattice. In the blue-shaded regime of Fig. \ref{fig2}(a), the system is in the nonlinear topologically trivial phase as described by the nonlinear Schr\"{o}dinger equation in Eqs. (\ref{GNLS}). The topological polarization in this regime takes the form $\bm{R}_{\rm T}(A) = 0$, indicating that there are no nonlinear topological boundary modes present at the edges of the lattice.

A remarkable feature of the system under consideration is its topological invariance under increasing mode amplitudes in both the non-trivial and trivial phases. This phenomenon arises from the fact that mode amplitudes have a global effect on the nonlinear dynamics of the system and do not alter the topological properties of the lattice. This explains the invariance of the topological Berry phases and polarization for growing mode amplitudes.

\subsection{Applications to the static modes in other nonlinear models}

To further explore the implications of the previous results for nonlinear systems, we investigate the static nonlinear modes in a different nonlinear model, namely the Lotka-Volterra network. To this end, we study the following nonlinear equations of motion with Lotka-Volterra-type nonlinear interactions, 
\begin{eqnarray}\label{LV}
 & {} & \partial_t \Psi^{(1)}_{\bm{r}} = \sum_{\bm{r}',i'=1,2}U(\Psi^{(1)}_{\bm{r}},(-1)^{i'+1}\Psi^{(i')}_{\bm{r}'}), \nonumber \\
 & {} & \partial_t \Psi^{(2)}_{\bm{r}} = \sum_{\bm{r}',i'=1,2}U(-\Psi^{(2)}_{\bm{r}},(-1)^{i'+1}\Psi^{(i')}_{\bm{r}'}).
\end{eqnarray}
Although the \emph{dynamical} properties between the Lotka-Volterra model and the Schr\"{o}dinger-type nonlinear equations in Eqs. (\ref{GNLS}) are significantly different~\cite{strogatz2018nonlinear}, their \emph{static} properties are remarkably similar, thanks to the presence of chiral symmetry. The reason is as follows. 


In the Schr\"{o}dinger equation, chiral symmetry ensures that the frequencies of nonlinear normal modes emerge in pairs of $\pm\omega$. Therefore, nonlinear topological mode must have zero-frequency, because if the frequency of the topological mode were non-zero, a chiral-partner topological mode with the frequency of $-\omega$ must appear. These two topological modes can interfere and break their topological nature. As a result, the nonlinear topological mode must be static. This static nonlinear topological mode is equivalent to setting $\mathrm{i}\partial_t\Psi_{\bm{r}}^{(i)}=0$ in the Schr\"{o}dinger equation. On the other hand, the stationary solutions in the Lotka-Volterra model are obtained by setting $\partial_t\Psi_{\bm{r}}^{(i)}=0$. This finding suggests that the (Schr\"{o}dinger) static mode is very much similar as the (Lotka-Volterra) static mode, because these two zero-frequency solutions are obtained by substituting $\mathrm{i}\partial_t\Psi_{\bm{r}}^{(i)}=0$ and $\partial_t\Psi_{\bm{r}}^{(i)}=0$ in the Schr\"{o}dinger and Lotka-Volterra models, respectively.

The notion of drawing a comparison between the static solution of the Lotka-Volterra and Schr\"{o}dinger models originates from seminal works~\cite{Lubensky2015phonons, Kane2014NP, Zhou2018PRL, VitelliPRX, Sun2020PRL, paulose2015topological, Zhou2019PRX, Chen2023IM}, which established an analogy between static mechanics in Newtonian equations of motion and the static solutions in chiral-symmetric Schr\"{o}dinger equations. The topological properties of Newtonian mechanics are defined by introducing an auxiliary chiral-symmetric Schr\"{o}dinger equation, where we can compute the topological index of this auxiliary Schr\"{o}dinger equation. We then utilize this auxiliary Schr\"{o}dinger topological number to describe the topology of static Newtonian mechanics, although the dynamical features of these two models differ significantly. 

Building upon this notion, the static solutions for both the Schr\"{o}dinger and Lotka-Volterra models can be obtained by setting $\partial_t\Psi^{(i)}_{\bm{r}}=0$ for $i=1,2$ and all $\bm{r}$. This analogy facilitates the derivation of the static nonlinear boundary mode in Eq. (\ref{LV}). The phase diagram of the Lotka-Volterra-type model is depicted by Fig. \ref{fig2}(a), comprising three distinct regions: the nonlinear gapless, topologically non-trivial, and trivial phases, represented by the grey, red, and blue-shaded areas, respectively.

The Lotka-Volterra-type model exhibits topologically distinct boundary properties in the red and blue areas of the phase diagram. In the red parameter region, the model exhibits the emergence of nonlinear static edge modes from the boundary of the square lattice. These edge modes are in line with the derived nonlinear topological polarization of $\bm{R}_{\rm T}(A) = \hat{e}_x$. It should be noted that the population of species must be real and positive numbers, which imposes a constraint on our analysis. Under this constraint, we obtain the nonlinear boundary mode 
\begin{eqnarray}\label{edgeMode2}
\left(\begin{array}{c}
\Psi^{(1)}_{\bm{r}}(t)\\
\Psi^{(2)}_{\bm{r}}(t)\\
\end{array}\right)
 = A(e^{-\kappa_+(k_y=0) n_x}-e^{-\kappa_-(k_y=0)n_x})
\left(\begin{array}{c}
1\\
1\\
\end{array}\right),\nonumber \\
\end{eqnarray}
where $A$ is the mode amplitude. The spatial decay rate satisfies $\kappa_\pm(k_y=0)>0$, and is explicitly expressed in Appendix B. This result indicates that the nonlinear boundary mode is exponentially localized on the left open boundary of the system, confirming its topological nature. The spatial decay rate of the nonlinear boundary mode is in perfect agreement with the decay properties of the topological edge modes in the linear and nonlinear regimes of the Schr\"{o}dinger-type equations in Eqs. (\ref{GNLS}). Moreover, the nonlinear boundary mode corresponds to the nonlinear topological polarization, $\bm{R}_{\rm T}(A)=\hat{e}_x$, in the topologically non-trivial phase. In contrast to the red region, the blue region of the phase diagram of the Lotka-Volterra-type model is devoid of nonlinear topological static modes. This behavior is in line with the derived trivial nonlinear topological polarization, $\bm{R}_{\rm T}(A) = 0$.

\begin{figure}[htbp]
\centering
\includegraphics[scale=0.58]{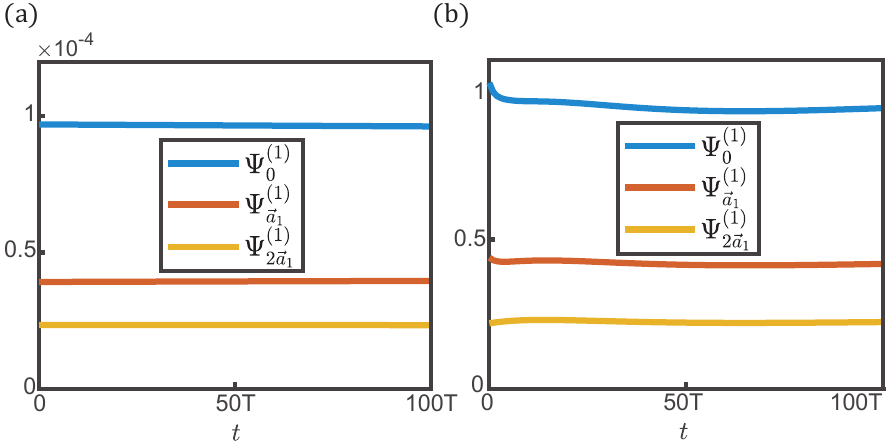}
\caption{The temporal evolutions of topologically protected boundary modes are shown in (a) and (b) for the weakly nonlinear and fully nonlinear regimes, respectively. In both cases, the modes evolve for $t=100T$ and still maintain their stability.
}\label{fig3}
\end{figure}

To validate the analytical results of the emergence of nonlinear topological boundary modes in the Lotka-Volterra-type model, we perform numerical simulations and present the results in Figs. \ref{fig2}(c,d) and Fig. \ref{fig3}. The numerical simulations confirm the emergence of the nonlinear topological modes on the open boundary that cuts the topological polarization vector $\bm{R}_{\rm T}(A)=\hat{e}_x$, as predicted by the analytical results. This topologically protected mode exhibits high stability in both temporal and spatial aspects, highlighting its robustness against small perturbations or fluctuations. We initialize the mode with $\pm10\%$ spatial fluctuations and $\pm10\%$ spatial variations in the nonlinear interactions, as shown by the bond lengths in Figs. \ref{fig2}(c,d). The numerical simulations indicate that the nonlinear topological mode remains highly stable even after $t=10^2T$ of self-oscillation, both in the weakly and strongly nonlinear regimes. The characteristic period of the nonlinear normal mode in the fully nonlinear regime is $T=2\pi$. The stability of the nonlinear topological mode confirms its robustness and reliability against small perturbations or fluctuations.


Chiral symmetry has notable effects on these nonlinear topological boundary modes. Firstly, this symmetry locks the frequencies of topological modes at zero, regardless of their amplitudes. These static modes are distinct from spatial symmetry-induced topological modes, which are sensitive to amplitudes and prone to losing nonlinear stability~\cite{Fronk2017JVA, Tempelman2021PRB}. Secondly, chiral symmetric nonlinear topological modes are unaffected by the breakdown of spatial symmetries, as observed in Figs. \ref{fig2}(c,d). Conversely, spatial symmetry-induced topological modes are quickly disrupted by spatial symmetry-breaking boundary conditions.


\section{Conclusions and outlook}

In this work, we extend the two non-spatial symmetries, including particle-hole symmetry and chiral symmetry, to nonlinear dynamics whose nonlinear wave functions do not necessarily possess $\textbf{U}(1)$-gauge symmetry. Chiral symmetry can quantize the Berry phase of nonlinear normal modes, determining their topological phases, and facilitating the emergence of nonlinear topological boundary modes. These modes have pinned frequencies at zero. They exhibit high stability against disruptions in spatial symmetries. Our work enables the non-spatial classification of nonlinear systems, expanding the ten-fold classification of linear topological insulators~\cite{Ryu2010NJP}. Our findings may suggest potential applications in nonlinear dynamics, where the stability conferred by nonlinear topology could aid in resilience against parameter changes.
\\

\emph{Note added---} Recently, we became aware of a related independent effort on the chiral symmetry of nonlinear dynamics~\cite{Fleury2024Arxiv}.

\section{Acknowledgment}

D. Z. would like to thank insightful discussions with Anthony J. Leggett, Xueda Wen, Junyi Zhang and Feng Li. D. Z. is supported by the National Natural Science Foundation of China (Grant Nos. 12374157, 12102039).

\appendix

\section{Berry phase of nonlinear normal modes}

In this section, we derive the adiabatic geometric phase of nonlinear normal modes. We consider the nonlinear equations of motion in a translationally invariant lattice, with $\mathrm{i}\partial_t\Psi(t) = H(\Psi)$, in which the plane-wave nonlinear normal mode reads $\Psi_{\bm{k},\omega}(\bm{k}\cdot\bm{r}-\omega t)$. We consider an adiabatic evolution to the nonlinear mode by slowly changing the wavevector $\bm{k}$ that follows a closed-loop trajectory in the Brillouin zone. Under the evolution, the nonlinear wave takes the format of 
\begin{eqnarray}\label{A0}
\Psi = \Psi_{\bm{k},\omega}\left(-\int^t \omega(t')dt'-\gamma(t)\right)
\end{eqnarray}
where $\gamma(t)$ is the adiabatic geometric phase that arises from the temporal evolution. This mode should satisfy the nonlinear equations of motion, which leads to 
\begin{eqnarray}\label{A1}
\mathrm{i}\partial_t \Psi & = & \mathrm{i}\left[\frac{\partial\Psi_{\bm{k},\omega}}{\partial\bm{k}}\frac{d\bm{k}}{dt}-\left(\omega+\frac{d\gamma}{dt}\right)\frac{\partial\Psi_{\bm{k},\omega}}{\partial\theta}+\partial_t\delta\Psi\right],\qquad 
\end{eqnarray}
where $\delta\Psi = \Psi-\Psi_{\bm{k},\omega}$ denotes mode change, and $\theta = \int^t\omega(t')dt'+\gamma(t)$ denotes the phase variable of the nonlinear mode. Since the shape of the nonlinear mode can vary during the process of adiabatic evolution, the nonlinear Hamiltonian also changes in this adiabatic process, leading to the expansion of the nonlinear Hamiltonian in terms of the mode variation $\delta\Psi$:
\begin{eqnarray}\label{A2}
H(\Psi) = 
H(\Psi_{\bm{k}}) +\left( \delta\Psi  \frac{\partial H}{\partial \Psi}+
\delta\Psi^* \frac{\partial H}{\partial \Psi^*}\right)_{\Psi_{\bm{k},\omega}}.
\end{eqnarray}
Employing the relationship $-\mathrm{i}\omega \partial_\theta \Psi_{\bm{k},\omega} = H(\Psi_{\bm{k},\omega})$, we combine Eqs. (\ref{A1}, \ref{A2}) and obtain the equations of motion for the nonlinear mode, which read 
\begin{eqnarray}\label{A3}
  & {} & \frac{\partial\Psi_{\bm{k},\omega}}{\partial\theta}\frac{d\gamma}{dt} = 
 \frac{\partial\Psi_{\bm{k},\omega}}{\partial\bm{k}}\frac{d\bm{k}}{dt}
 +\nonumber \\
  & {} & \mathrm{i}\left(\delta\Psi\frac{\partial H}{\partial\Psi}+\delta\Psi^*\frac{\partial H}{\partial \Psi^*}-\mathrm{i}\partial_t\delta\Psi\right)_{\Psi_{\bm{k},\omega}}.
\end{eqnarray}
Eq. (\ref{A3}) contains two terms on the right-hand side. The first term corresponds to the Berry phase of nonlinear normal modes in nonlinear normal modes. The second term arises from the change in the mode shape and amplitude due to the adiabatic evolution, which can affect the nonlinear Hamiltonian of the system. Together, these terms contribute to the overall adiabatic geometric phase of the nonlinear normal modes.

To gain a more detailed understanding of the nonlinear normal mode, we express it in terms of a Fourier series. Specifically, we write $\Psi_{\bm{k},\omega}(\theta) = \sum_l (\psi_{l\bm{k}}^{(1)}, \psi_{l\bm{k}}^{(2)}e^{\mathrm{i}l\phi_{\bm{k}}})^\top e^{\mathrm{i}l\theta}$, where $\psi_{l\bm{k}}^{(1)}$ and $\psi_{l\bm{k}}^{(2)}$ are the Fourier components of the mode in the two wave components, respectively, and $\phi_{\bm{k}}$ is the relative phase between the two components. Thus, we obtain
\begin{eqnarray}\label{A4}
 & {} & \frac{\partial\Psi_{\bm{k},\omega}}{\partial\theta} = \sum_l \left(\psi_{l\bm{k}}^{(1)},\psi_{l\bm{k}}^{(2)}e^{\mathrm{i}l\phi_{\bm{k}}}\right)^\top \mathrm{i}l e^{\mathrm{i}l\theta},\nonumber \\
 & {} & \frac{\partial\Psi_{\bm{k},\omega}}{\partial\bm{k}} = \sum_l \left[\frac{\partial\psi_{l\bm{k}}^{(1)}}{\partial\bm{k}},\left(\frac{\partial\psi_{l\bm{k}}^{(2)}}{\partial\bm{k}}+\mathrm{i}l\psi_{l\bm{k}}^{(2)}\frac{\partial\phi_{\bm{k}}}{\partial\bm{k}}\right)e^{\mathrm{i}l\phi_{\bm{k}}}\right]^\top e^{\mathrm{i}l\theta}.\nonumber \\
\end{eqnarray}
We now use Eqs. (\ref{A4}) to substitute results into Eq. (\ref{A3}). We then multiply both sides of Eq. (\ref{A3}) by $\Psi^\dag_{\bm{k},\omega}$ and integrate over the phase variable $\theta$ from $0$ to $2\pi$. This procedure leads us to derive the following relationship:
\begin{eqnarray}\label{A5}
 & {} & \frac{d\gamma}{dt}\sum_l l\sum_{j} |\psi_{l\bm{k}}^{(i)}|^2
 = \nonumber \\
 & {} & \frac{d\bm{k}}{dt}\sum_l \left(l\frac{\partial\phi_{\bm{k}}}{\partial\bm{k}}|\psi_{l\bm{k}}^{(2)}|^2-\mathrm{i}\sum_j \psi_{l\bm{k}}^{(j)*}\frac{\partial\psi_{l\bm{k}}^{(i)}}{\partial \bm{k}}\right)
+\nonumber \\
 & {} & \int_0^{2\pi}\frac{d\theta}{2\pi}\Psi_{\bm{k},\omega}^\dag\left(\delta\Psi\frac{\partial H}{\partial\Psi}+\delta\Psi^*\frac{\partial H}{\partial \Psi^*}- H(\Psi)\delta\Psi\right)_{\Psi=\Psi_{\bm{k},\omega}}\quad
\end{eqnarray}
where the relationship $\mathrm{i}\partial_t\delta\Psi = H(\Psi)\delta\Psi$ has been adopted. Finally, we integrate over the time variable $t$ and obtain the Berry phase of nonlinear normal modes:
\begin{eqnarray}\label{A6}
\gamma_C = \gamma_C^{(\rm B)}+\gamma_C^{(\rm NL)}.
\end{eqnarray}
In Eq. (\ref{A6}), there are two contributions in the adiabatic geometric phase. The first term is the Berry phase of nonlinear normal modes, 
\begin{eqnarray}\label{A7}
\gamma_C^{(\rm B)}
 = \oint_C d\bm{k}\frac{\sum_l \left(l\frac{\partial\phi_{\bm{k}}}{\partial\bm{k}}|\psi_{l\bm{k}}^{(2)}|^2-\mathrm{i}\sum_j \psi_{l\bm{k}}^{(j)*}\frac{\partial\psi_{l\bm{k}}^{(i)}}{\partial \bm{k}}\right)}{\sum_l l\sum_{j} |\psi_{l\bm{k}}^{(i)}|^2}.\quad
\end{eqnarray}
The second term 
\begin{eqnarray}\label{A8}
 & {} & \gamma_C^{(\rm NL)}
 = \int dt\bigg(\sum_l l\sum_{j} |\psi_{l\bm{k}}^{(i)}|^2\bigg)^{-1}\nonumber \\
 & {} & \int_0^{2\pi}\frac{d\theta}{2\pi}\Psi_{\bm{k},\omega}^\dag\left(\delta\Psi\frac{\partial H}{\partial\Psi}+\delta\Psi^*\frac{\partial H}{\partial \Psi^*}-H(\Psi)\delta\Psi\right)_{\Psi=\Psi_{\bm{k},\omega}}\nonumber \\
\end{eqnarray}
is known as the ``purely nonlinear" contribution to the adiabatic geometric phase, as it arises solely from the nonlinear interactions within the system. Specifically, it represents the change in the nonlinear Hamiltonian with respect to the variation of the nonlinear normal mode, which arises due to the dependence of the Hamiltonian on the mode amplitude. We emphasize that this term vanishes for the purely linear Sch\"{o}dinger-type equations.

To investigate the relationship between the conventional linear Berry phase and Berry phase of nonlinear normal modes, we consider the Hamiltonian for the linear and nonlinear Schr\"{o}dinger equations that respect the $\textbf{U(1)}$-gauge symmetry in the wave functions, which is given by 
\begin{eqnarray}\label{A10}
H(\Psi) = H_0\Psi+g|\Psi|^2 \Psi.
\end{eqnarray}
Here, $H_0$ is the linear part of the dynamics, represented by a matrix, and $g$ is the coefficient of the nonlinearity. In these systems, the nonlinear normal modes can be represented by sinusoidal waves,
\begin{eqnarray}\label{A9}
\Psi_{\bm{k},\omega}(\bm{k}\cdot\bm{r}-\omega t) & = &
\Psi_{\bm{k}} e^{\mathrm{i}(\bm{k}\cdot\bm{r}-\omega t)}\nonumber \\
 & = & 
(\Psi_{\bm{k}}^{(1)}, \Psi_{\bm{k}}^{(2)}e^{\mathrm{i}\phi_{\bm{k}}})^\top e^{\mathrm{i}(\bm{k}\cdot\bm{r}-\omega t)}. \qquad
\end{eqnarray}
Here, the normalization condition given by $|\Psi_{\bm{k}}^{(1)}|^2+|\Psi_{\bm{k}}^{(2)}|^2=1$ has been adopted.

To simplify the Berry phase of nonlinear normal modes, $\gamma_C^{(\rm B)}$, and the purely nonlinear contribution to the adiabatic geometric phase, $\gamma_C^{(\rm NL)}$, we employ the relationships given by the representation of nonlinear normal modes as sinusoidal waves, as presented in Eq. (\ref{A9}). In particular, we note that nonlinear plane waves in these systems contain only the fundamental harmonic, which allows us to simplify the expressions for $\psi_{l\bm{k}}^{(1)}$ and $\psi_{l\bm{k}}^{(2)}$ as $\psi_{l\bm{k}}^{(1)} = \Psi^{(1)}_{\bm{k}}\delta_{l1}$ and $\psi_{l\bm{k}}^{(2)} = \Psi^{(2)}_{\bm{k}}\delta_{l1}$, respectively. Plugging these simplifications into $\gamma_C^{(\rm B)}$ reduces the Berry phase of nonlinear normal modes as 
\begin{eqnarray}\label{A11}
\gamma_C^{(\rm B)}
 & = & 
\oint_{C}d\bm{k}\cdot \frac{\sum_{l} \left( l |\psi_{l\bm{k}}^{(2)} |^2 \frac{\partial\phi_{\bm{k}}}{\partial \bm{k}}+\mathrm{i}\sum_j\psi_{l\bm{k}}^{(j)*}\frac{\partial\psi_{l\bm{k}}^{(i)}}{\partial\bm{k}}\right)\delta_{l1}}
{\sum_{l'}  l'  \left(\sum_{j'}|\psi_{l'\bm{k}}^{(j')}|^2\right)\delta_{l'1}}\nonumber \\
 & = & \oint_{C}d\bm{k}\cdot \left( |\psi_{1\bm{k}}^{(2)} |^2 \frac{\partial\phi_{\bm{k}}}{\partial \bm{k}}+\mathrm{i}\sum_j\psi_{1\bm{k}}^{(j)*}\frac{\partial\psi_{1\bm{k}}^{(i)}}{\partial \bm{k}}\right) \nonumber \\
 & = & 
 \oint_{C}d\bm{k}\cdot\mathrm{i}\left(\sum_j \Psi_{\bm{k}}^{(j)*}\partial_{\bm{k}}\Psi_{\bm{k}}^{(i)}-\mathrm{i}|\Psi_{\bm{k}}^{(2)}|^2\partial_{\bm{k}}\phi_{\bm{k}}\right) \nonumber \\
 & = &
 \oint_{C}d\bm{k}\cdot\mathrm{i} (\Psi_{\bm{k}}^{(1)*},\Psi_{\bm{k}}^{(2)*}e^{-\mathrm{i}\phi_{\bm{k}}}) \partial_{\bm{k}} \left(
 \begin{array}{c}
 \Psi_{\bm{k}}^{(1)}\\
 \Psi_{\bm{k}}^{(2)}e^{\mathrm{i}\phi_{\bm{k}}}\\
 \end{array}
 \right)\nonumber \\
 & = & 
 \oint_{C}d{\bm{k}}\cdot\mathrm{i}\langle \Psi_{\bm{k}}|\partial_{\bm{k}}|\Psi_{\bm{k}}\rangle = \gamma_{C,{\rm linear}}^{(\rm B)},
\end{eqnarray}
where $\Psi_{\bm{k}} = (\Psi_{\bm{k}}^{(1)},\Psi_{\bm{k}}^{(2)}e^{\mathrm{i}\phi_{\bm{k}}})^\top$ is the eigenvector of the Hamiltonian, and $\gamma_{C,{\rm linear}}^{(\rm B)}$ denotes the conventional form of Berry phase in linear systems. At this point, we have shown that the Berry phase of nonlinear normal modes, as expressed in Eq. (\ref{gamma1}), can be reduced to the conventional Berry phase when the nonlinear wave functions yield $\textbf{U}(1)$-gauge symmetry.

When considering nonlinear interactions of the form 
\begin{eqnarray}\label{A12}
H(\Psi) = H_0\Psi+g|\Psi|^2 \Psi,
\end{eqnarray} 
we can compute the contribution of nonlinear interactions to the adiabatic geometric phase, $\gamma_C^{(\rm NL)}$. By using the relationships $\partial H(\Psi)/\partial \Psi = H_0+2g|\Psi|^2$ and $\partial H(\Psi)/\partial\Psi^* = g\Psi^2$, we can substitute into $\gamma_C^{(\rm NL)}$ to obtain the result 
\begin{eqnarray}\label{A14}
& {} & \gamma_C^{(\rm NL)}\nonumber \\
 & = &\int dt \bigg(\sum_l l\sum_{j} |\psi_{l\bm{k}}^{(i)}|^2\delta_{l1}\bigg)^{-1}\nonumber \\
 & {} & \int_0^{2\pi}\frac{d\theta}{2\pi}\Psi_{\bm{k},\omega}^\dag\left(\delta\Psi\frac{\partial H}{\partial\Psi}+\delta\Psi^*\frac{\partial H}{\partial \Psi^*}-H(\Psi)\delta\Psi\right)_{\Psi_{\bm{k},\omega}}\nonumber \\
 & = &\int dt \int_0^{2\pi}\frac{d\theta}{2\pi}\nonumber \\
 & {} & \Psi_{\bm{k},\omega}^\dag\left(\delta\Psi\frac{\partial H}{\partial\Psi}+\delta\Psi^*\frac{\partial H}{\partial \Psi^*}-H(\Psi)\delta\Psi\right)_{\Psi_{\bm{k},\omega}}\nonumber \\
 & = & \int dt\,\, g\left(\Psi^2\Psi^{\dag}  \delta\Psi^\dag+\Psi^{\dag2}\Psi\delta\Psi\right)_{\Psi_{\bm{k},\omega}}.
\end{eqnarray}

\section{Analytical results of nonlinear topological boundary modes}

In this section, we analyze the decay rate of the nonlinear boundary mode by setting the static condition, $\partial\Psi_{\bm{r}}^{(i)}/\partial t=0$, and simplifying the Schr\"{o}dinger-type nonlinear equations of motion, as given by 
\begin{eqnarray}\label{B1}
 & {} & \Psi^{(1)}_{\bm{r}}-\sum_{j=x,y}g_j (\Psi^{(1)}_{\bm{r}+\hat{e}_j}+\Psi^{(1)}_{\bm{r}-\hat{e}_j}) -\Psi^{(2)}_{\bm{r}+\hat{e}_x}+  \Psi^{(2)}_{\bm{r}-\hat{e}_x}=0, \nonumber \\
 & {} & \Psi^{(2)}_{\bm{r}}-\sum_{j=x,y}g_j(\Psi^{(2)}_{\bm{r}+\hat{e}_j}+\Psi^{(2)}_{\bm{r}-\hat{e}_j})-\Psi^{(1)}_{\bm{r}+\hat{e}_x}+\Psi^{(1)}_{\bm{r}-\hat{e}_x}=0.\nonumber \\
\end{eqnarray}
We analyze the behavior of the nonlinear boundary mode, which is exponentially localized on the left open boundary and takes the form of a plane wave in the transverse $y$ direction. Using this ansatz, we derive the wave amplitudes and substitute from the static motion equations given by Eq. (\ref{B1}). The resulting solution, as given by Eq. (\ref{edgeMode1}), provides an analytic expression for the spatial decay rates of the edge mode, which are expressed as 
\begin{eqnarray}\label{drate2}
 & {} & \kappa_\pm(k_y) = \nonumber \\
 & {} & -\ln\bigg[\frac{\frac{1}{2}-g_y\cos k_y\pm\sqrt{(\frac{1}{2}-g_y\cos k_y)^2+1-g_x^2}}{g_x+1}\bigg].\qquad
\end{eqnarray}
The analytical expression for the spatial decay rates of the nonlinear boundary mode, as given by Eq. (\ref{drate2}), reveals important insights into the behavior of the mode at different values of the nonlinearity parameters and the transverse wavevector. In particular, we find that within the red region of parameter space in Fig. \ref{fig2}(a), where $-g_x+1/2<g_y<g_x-1/2 \bigcup g_x+1/2<g_y<-g_x-1/2$, the decay rates are positive, indicating that the mode described by Eq. (\ref{edgeMode1}) is exponentially localized on the left open boundary of the square lattice.


%

\end{document}